\documentclass[%
 reprint,
 amsmath,amssymb,
 aps,
]{revtex4-1}

\usepackage{graphicx}
\usepackage{dcolumn}
\usepackage{bm}

\usepackage{here}
\usepackage{comment}
\usepackage{color}

\begin{document}


\title{Superdiffusion of Aerosols Emitted After Sneezing \\ - Nonequilibrium Statistical Mechanics Approach -}

\author{Satoshi SUGIMOTO}
\email{alohngoabnp@gmail.com}
 \affiliation{Deparment of Applied Mathematics and Physics, Graduate School of Informatics, Kyoto University}
\author{Ken UMENO}%
 \email{umeno.ken.8z@kyoto-u.ac.jp}
\affiliation{Deparment of Applied Mathematics and Physics,  Graduate School of Informatics, Kyoto University
}%



\date{\today}

\begin{abstract}

We study a stochastic behavior of aerosols by the non-equilibrium statistical mechanics approach using the analytical approach of the Langevin equation.   
We firstly show that superdiffusion can possibly occur right after the emission, which may be attributed to the physical mechanism of the outbreak of the COVID-19 pandemic. We also provide clear evidence of the least required distance to prevent infections occurred by aerosols. In particular, the required distance to prevent aerosol infections is derived to be about 42 m when we assume Cauchy distribution as an initial velocity distribution. This fact implies that due to superdiffusion the aerosol infection can occur even far away from a long distance as compared to the previous considerations.
\end{abstract}


\maketitle


\section{Introduction}
Diffusion is one of the important issues in non-equilibrium statistical mechanics, such as the behavior of a drop of ink spreading into water. This phenomenon is observed not only in physical phenomena but also in biological phenomena. Okubo and Umeno studied the phenomenon of diffusion in dynamical systems \cite{okubokennichi}. They showed that in systems with a Hamiltonian with time-reversibility, an irreversible behavior occurs in the course of time. Alexander analyzed the invasion of cancer cells in tissues using the diffusion equation with real derivatives, and observed {\it superdiffusion} \cite{cancer}. P. Siegle {\it et.al} numerically studied a minimal non-Markovian model of superdiffusion which originates from long-range velocity correlations within the generalized Langevin equation approach \cite{Goychuk}.

In this Letter, we investigate the diffusion phenomenon of aerosols. Aerosols are many particles dispersed in gas, discussed in different fields of variant researches, such as environmental problems, air purification, medical applications, and so on \cite{aerobook}. These studies are mostly based on experimental or theoretical approaches and simulation approaches of fluid dynamics. One of the most general issues on aerosol dynamics is droplets generated by sneezes. Silwal {\it et.al} determine how many aerosols are dispersed by each type of pronunciation \cite{sound}. Bourouiba {\it et.al} photographed the behavior of the aerosols generated by a sneeze \cite{Bourouiba}. Fluid dynamics simulations are used to investigate the difference in the degree of aerosol diffusion with and without a mask \cite{mask}. However, aerosol infection cannot be explained without simulation and there are inevitable simulation errors due to its complex fluid phenomena such as turbulent behavior even for small time duration such as several seconds. Alternatively, we discuss the diffusive behavior of aerosols from the viewpoint of stochastic processes, which is different from the previous studies. Our findings provide an evidence that the degree of diffusion of aerosols well describes the actual phenomena, and show that an approach based on stochastic processes is also effective for analyzing aerosol dynamics. Furthermore, if the initial distribution of velocity is a Cauchy distribution, it is theoretically expected that {\it superdiffusion} occurs right after a sneeze. Note that  {\it superdiffusion} of stochastic processes is not superdiffusion due to turbulent fields of fluid dynamics. In this paper, we suppose that the behavior of an aerosol after a sneeze follows the Langevin equation. We ignore the effect of turbulent field and flows of air for describing diffusive behavior after a sneeze. We can thus obtain the distribution of their positions from the Langevin equation for their velocity \cite{kubobook} where the turbulent behavior 
at the moment of sneezing is used only for the determination of the initial velocity distribution obeying Cauchy distribution.

\section{Calculating the Distribution of Position Using Langevin Equation of Velocity}
Langevin equation for velocity of a particle of mass $m$ is 
\[  \displaystyle
\dot{u}(t)=-\gamma{u}+R(t)/m
\]
where $\gamma$ is a friction coefficient of a particle and $R(t)$ is a white noise, $\langle{R(t_{1})R(t_{2})}\rangle=2m\gamma kT\delta(t_1-t_2)$.  Here, we consider a white noise $R(t)$ to describe the source of random fluctuations of aerosols. When the initial distribution of a velocity is a normal distribution with its average $U$ and standard deviation $\displaystyle V_{0}=\frac{\tan10^{\circ}}{1-\tan10^{\circ}}U$ \cite{airborne}, the result is 
\begin{equation}
\displaystyle 
P(x,t)=\mathcal{N}_{x}\left(U\frac{1-e^{-\gamma t}}{\gamma},2\sigma^{2}+V^{2}_{0}\left(\frac{1-e^{-\gamma t}}{\gamma}\right)^{2}\right),
\label{Gauss}
\end{equation}
where $\mathcal{N}_{x}(\mu,\sigma^{2})$ represents a normal distribution with a mean $\mu$ and a variance $\sigma^{2}$ with respect to $x$, and $\sigma^{2}$ is defined as $\displaystyle
\frac{kT}{m\gamma}\left(t-\frac{2}{\gamma}(1-e^{-\gamma t})+\frac{1}{2\gamma}(1-e^{-2\gamma t})\right)$ \cite{kampen}. To examine the distribution of positions, we calculate the characteristic functional of $x(t)$ 

\[
\displaystyle
\footnotesize
\begin{split}
&\langle e^{i\xi{x(t)}} \rangle\\
&=\left\langle \exp\left(i\xi{u_{0}}\frac{1-e^{-\gamma t}}{\gamma}\right)\right\rangle \left\langle\exp\left(i\xi \int^{t}_{0}d\tau\frac{1-e^{-\gamma(t-\tau)}}{\gamma}\frac{R(\tau)}{m}\right)\right\rangle,
\end{split}
\]
where $\langle A\rangle$ means an expected value of A over all the possible stochastic processes $x(t)$ computed by a stochastic integral. The calculation of the first factor is as follows:

\[  \displaystyle
\begin{split}
&\left\langle \exp\left(i\xi{u_{0}}\frac{1-e^{-\gamma t}}{\gamma}\right)\right\rangle\\
&=\int\exp\left(i\xi{u_{0}}\frac{1-e^{-\gamma t}}{\gamma}\right)p(u_{0})du_{0}\\
&=\mathcal{F}[p](\tilde{\xi})
\end{split}
\]
where $\displaystyle \tilde{\xi}=\frac{1-e^{-\gamma t}}{\gamma}\xi$ and $\displaystyle \mathcal{F}[p](\xi) = \int^{\infty}_{-\infty} p(x)e^{i\xi x}dx$. Then $\langle e^{i\xi{x(t)}} \rangle$ can be calculated as:
\[  \displaystyle
\langle e^{i\xi{x(t)}} \rangle=\exp(-\xi^{2}\sigma^{2})\mathcal{F}[p](\tilde{\xi}),
\]

\[  \displaystyle
P(x,t)= \frac{1}{2\pi}\int^{\infty}_{-\infty}\exp( -\sigma^{2}\xi^{2}-i\xi x)\mathcal{F}[p](\tilde{\xi})d\xi.
\]
 We assume that value of $\sigma^{2}$ is small. We analyze this integral by using the Taylor expansion \cite{landau}:

\begin{equation}
\displaystyle
\small
\begin{split}
&=\sum_{n=0}^{\infty}\frac{1}{2\pi}\int^{\infty}_{-\infty}(-1)^{n}\frac{\sigma^{2n}\xi^{2n}}{n!} \exp(- i\xi x)\mathcal{F}[p](\tilde{\xi})d\xi\\ 
&=\sum_{n=0}^{\infty}\frac{1}{2\pi}\frac{\partial^{2n}}{\partial x^{2n}}\int^{\infty}_{-\infty}\frac{\sigma^{2n}}{n!} \exp(- i\xi x)\mathcal{F}[p](\tilde{\xi})d\xi\\ 
&=\sum_{n=0}^{\infty}\frac{\gamma}{1-e^{-\gamma t}}\frac{\sigma^{2n}}{2\pi n!}\frac{\partial^{2n}}{\partial x^{2n}}\int^{\infty}_{-\infty} \exp\left(-i\tilde{\xi} \frac{\gamma x}{1-e^{-\gamma t}}\right)\mathcal{F}[p](\tilde{\xi})d\tilde{\xi}\\ 
&=\frac{\gamma}{1-e^{-\gamma t}}\sum_{n=0}^{\infty}\frac{\sigma^{2n}}{n!}\frac{\partial^{2n}}{\partial x^{2n}}p\left(\frac{\gamma x}{1-e^{-\gamma t}}\right)\\ 
\end{split}
\label{conclusion}
\end{equation}
In particular, when $\sigma^{2}$ is evaluated as zero, $P(x,t)$ is wrtitten as
\[
P(x,t)=\frac{\gamma}{1-e^{-\gamma t}}p\left(\frac{\gamma x}{1-e^{-\gamma t}}\right).
\]

Now we assume that the initial velocity $u_{0}=u(0)$ follows Cauchy distribution $\displaystyle \frac{1}{\pi}\frac{\Gamma}{(u_{0}-U)^{2}+\Gamma^{2}}$, where $\Gamma$ is the scaling parameter and satisfies $\Gamma=V_{0}$. The reasons of the choice of the Cauchy distribution as the initial velocity are as follows. 1:  Tong and Goldburg \cite{Tong}  showed by the experiment of photon correlation spectroscopy that 
  for the velocity difference distribution of the turbulent flow in the case of small separation, in both two and three dimensions, a Cauchy distribution obeys. 2: Such
 experimental results on the Cauchy distribution  have also theoretical explanations as Onuki \cite {Onuki} and Min et al. \cite{Min}. 3: We can safely assume that {\it only at the moment of  sneezing,}
  the initial velocity distribution of aerosols obeys the velocity difference (Cauchy) distribution of the turbulent flow.
Furthermore, the Cauchy distribution is one representative of the class of L\'{e}vy stable distributions including a normal distribution as a special distribution\cite{Levy} and can be categorized as a limiting distribution of the generalized central limit theorem which is important in statistical physics \cite{Gnedenko}. Using the Taylor expansion up to the second order term in Eq. (2),  we get the distribution of aerosol $P(x,t)$:
\begin{equation}
\displaystyle
P(x,t) \approx \frac{1}{\pi}\frac{\tilde{\Gamma}}{\tilde{x}^{2}+\tilde{\Gamma}^{2}}\left(1-2\sigma^{2}\frac{\tilde{\Gamma}^{2}-3\tilde{x}^{2}}{(\tilde{x}^{2}+\tilde{\Gamma}^{2})^{2}}\right),
\label{cauchytaylor}
\end{equation}
where $\displaystyle \tilde{\Gamma} = \frac{1-e^{-\gamma t}}{\gamma}\Gamma$ and $\displaystyle \tilde{x} = x-\frac{1-e^{-\gamma t}}{\gamma}U$. Note that $\displaystyle \int^{\infty}_{-\infty}\frac{1}{\pi}\frac{\tilde{\Gamma}}{\tilde{x}^{2}+\tilde{\Gamma}^{2}}\left(-2\sigma^{2}\frac{\tilde{\Gamma}^{2}-3\tilde{x}^{2}}{(\tilde{x}^{2}+\tilde{\Gamma}^{2})^{2}}\right)dx = 0$ showing that \(P(x,t)\) in Eq. (3) is exactly 
a probability density function. 
 \color{black} In particular, when $\sigma^{2} \ll 1$ such that we can ignore the second term, we can thus derive the Cauchy distribution. This formula implies the deceleration against the time.  The properties of the distribution are  also preserved even after this approximation with higher-order expansions as long as  $\displaystyle \frac{\sigma^{2}}{\Gamma^{2}}<<1$. In particular, when we assume $\displaystyle t<<\frac{1}{\gamma}$,
\[  \displaystyle
P(x,t)\approx\frac{1}{\pi}\frac{\Gamma{t}}{(\Gamma{t})^{2}+(x-Ut)^{2}}
\]
holds, which show that $x(t)$ obeys a Cauchy distribution centered at $x=Ut$ with scale number $\Gamma{t}$. This result indicates the occurrence of {\it superdiffusion} and the mean-aquare displacement (MSD) obeys to the order of $O(t^2)$, where the order of diffusion speed is much 
faster than the normal diffusion having the speed as $\mathrm{MSD} = O(t)$.


\section{Numerical Estimation}
We numerically estimate a required distance to prevent aerosol infection. The properties in numerical estimation are as follows: 
\begin{enumerate}
\item The density of the aerosol is assumed to be the same as that of water.
\item Temperature is set to the typical temperature as $291\mathrm{K}$ \cite{aeromove}.
\item $U$ is assumed 50~m/s \cite{aeromove}.
\item  The Friction Coefficient of Aerosols $\gamma$

In this paper we ignore the effect of turbulent field and flows of air \textit{after sneezing} , so the Stokes formula \cite{aerobook} is valid during the diffusive process in this case. This hypothesis implies that the friction coefficient of aerosols $\gamma$ is $\displaystyle \frac{3\pi\mu d_{p}}{m}$. Note that $\mu$ is the viscosity coefficient of gas and $d_p$ is the diameter of the aerosol. The mass $m$ is calculated as $\frac{1000}{6} \pi d_{p}^{3}~\mathrm{kg}$ at this time.
We consider aerosol particle size and the corresponding $\gamma$ in the TABLE \ref{gamma}.
\begin{table}[H]
\begin{center}
\caption{Aerosol particle size and $\gamma$}
\begin{tabular}{c|c}
\hline
Droplet diameter $\mathrm{\mu{m}}$&$\gamma~\mathrm{s^{-1}}$\\ \hline
~60&$90.0$\\
~80&$50.6$\\
100&$32.4$\\
120&$22.5$\\
140&$16.5$\\
200&$~8.1$\\
\hline
\end{tabular}
\label{gamma}
\end{center}
\end{table}

\item Aerosol Residence Time $t$

TABLE \ref{time} shows aerosol residence time at different relative humidity \cite{aeromove}.
\begin{table}[H]
\begin{center}
\caption{Aerosol residence time $t~\mathrm{s}$ at different relative humidity}
\begin{tabular}{c||c|c|c|c|c}
\hline
$d_{p}~\mathrm{\mu{m}}$&$0\%$&$30\%$&$50\%$&$70\%$&$90\%$ \\ \hline
$~60$&~3&~4&12&~7&35\\
$~80$&~5&~7&18&12&13\\
$100$&~8&12&12&17&~9\\
$120$&10&~8&~6&~7&~6\\
$140$&~7&~6&~6&~6&~6\\
$200$&~5&~5&~5&~5&~5\\
\end{tabular}
\label{time}
\end{center}
\end{table}

\item The mean free path of the aerosol $l_{p}$\\
The mean free path corresponds to the expected distance that a particle travels between two collisions \cite{atkins}. Collisions between aerosol particles are negligible. We assume that the volume of exhaled air is $0.001\,\mathrm{m^{3}}$ \cite{physiology}. Since the mean free path cannot be uniquely determined for a non-equilibrium system, we calculate the velocity and particle size distribution of aerosols at the moment of sneezing. The result of the calculation is shown in TABLE \ref{meanfreepath}.

\begin{table}[H]
\begin{center}
\caption{Aerosol particle diameter $d_{p}$ $\mathrm{\mu{m}}$ and the mean free path $l_{p}~\mathrm{m}$ due to collisions of air constituent molecules. The notation of the numbers is in significant figures, e.g. 6.3e-3 represents $6.3\times10^{-3}$.}
\begin{tabular}{c|c}
\hline
$d_{p}~\mathrm{\mu{m}}$&$l_{p}~\mathrm{m}$ \\ \hline
$~60$&$2.9\times10^{-18}$\\
$~80$&$1.6\times10^{-18}$\\
$100$&$1.0\times10^{-18}$\\
$120$&$7.3\times10^{-19}$\\
$140$&$5.4\times10^{-19}$\\
$200$&$2.6\times10^{-19}$\\
\hline
\end{tabular}
\label{meanfreepath}
\end{center}
\end{table}

\item When an initial velocity $u_{0}=u(0)$ follows Cauchy distribution, we can evaluate the second term of (\ref{cauchytaylor}) as follows:

\[
\begin{split}
\displaystyle
\left|2\sigma^{2}\frac{\tilde{\Gamma}^{2}-3\tilde{x}^{2}}{(\tilde{x}^{2}+\tilde{\Gamma}^{2})^{2}}\right|&\leq 6\frac{\sigma^{2}}{\tilde{x}^{2}+\tilde{\Gamma}^{2}}\\
&\leq\frac{6\sigma^{2}}{\tilde{\Gamma}^{2}}\\
&=\frac{6\gamma kT}{m\Gamma^{2}}\frac{\int^{t}_{0}\left(\frac{1-e^{-\gamma \tau}}{\gamma}\right)^{2}d\tau}{\left(\frac{1-e^{-\gamma t}}{\gamma}\right)^{2}}\\
&\leq\frac{6\gamma kTt}{m\Gamma^{2}}.
\end{split}
\]
When the diameter of an particle $d_{p}$ is $200\mathrm{\mu m}$, we have $\displaystyle \frac{\gamma kTt}{m\Gamma^{2}}=6.8\times10^{-14}t$. This means that the second term of Eq. (\ref{cauchytaylor}) is negligible in this numerical estimation.

\end{enumerate}

TABLE \ref{gausstable1} shows the numerical results in the condition that the distribution for an initial velocity $u_{0}$ is the normal distribution with its average $U$ and the standard deviation $V_{0}$. The time $t$ is calculated as the mean and the variance of the aerosols at the time listed in the previous TABLE \ref{time}.


\begin{table}[H]
\begin{center}
\caption{The mean and the standard deviation in meters of the distribution of aerosols at each particle size $d_{p}$. The effect of relative humidity is negligible.}

\begin{tabular}{c||c|c}
\hline
$d_{p}~\mathrm{\mu{m}}$&Mean m&Deviation m \\ \hline
$~60$&0.56&0.119\\
$~80$&0.99&0.211\\
$100$&1.54&0.330\\
$120$&2.22&0.476\\
$140$&3.03&0.648\\
$200$&6.17&1.321\\
\end{tabular}
\label{gausstable1}
\end{center}
\end{table}

TABLE \ref{cauchytable1} shows the numerical results in the condition that the distribution for an initial velocity $u_{0}$ is the Cauchy distribution. The time $t$ is calculated as the mean and the variance of the aerosols at the time listed in the previous Table \ref{time}.

\begin{table}[H]
\begin{center}
\caption{$\displaystyle U\frac{1-e^{-\gamma t}}{\gamma}$ and the scale number $\tilde{\Gamma}$ at each particle size $d_{p}$ when the initial velocity $u_{0}$ follows the Cauchy distribution. The effect of relative humidity is negligible.}
\begin{tabular}{c||c|c}
\hline
$d_{p}~\mathrm{\mu{m}}$&$\displaystyle U\frac{1-e^{-\gamma t}}{\gamma}$ m&$\tilde{\Gamma}$ m\\ \hline
$~60$&0.56&0.119\\
$~80$&0.99&0.211\\
$100$&1.54&0.330\\
$120$&2.22&0.476\\
$140$&3.03&0.648\\
$200$&6.17&1.321\\
\end{tabular}
\label{cauchytable1}
\end{center}
\end{table}

\begin{figure}[H]
\begin{center}
\includegraphics[width=80mm]{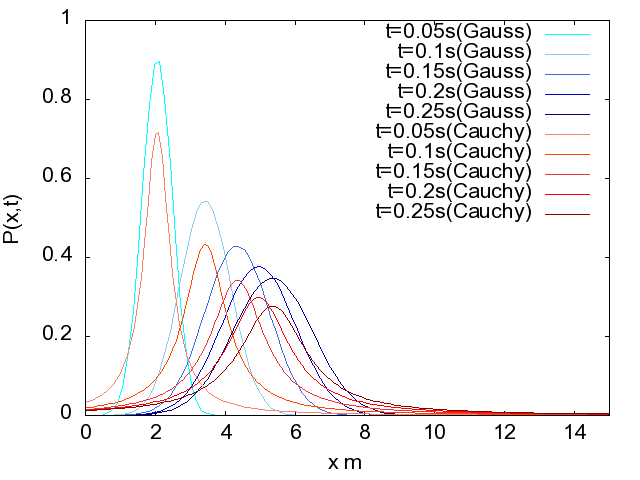}
\caption{The results of how droplets diffuse in one sneeze ($~\Gamma=\frac{50~\tan 10^{\circ}}{1-\tan 10^{\circ}}~\mathrm{m/s}$). $d_{p}=200~\mathrm{\mu m}$ is used.}
\label{cauchy}
\end{center}
\end{figure}

FIG \ref{cauchy} shows how droplets diffuse calculated from Eqs. (\ref{Gauss}) and (\ref{cauchytaylor}). The result indicates that the maximum point moves forward as the mount of time increases.

We evaluate a required distance $\alpha$ to prevent aerosol infection. $\alpha$ is defined as the distance for that $1\%$ of aerosols diffuse: $\displaystyle \int^{\infty}_{\alpha}P(x,t)dx=0.01$.
When $P(x,t)$ follows a normal distribution $P(x,t)=\mathcal{N}_{x}(\mu,\sigma^{2})$, we have

\[
\displaystyle 
\begin{split}
\int^{\infty}_{\alpha}\mathcal{N}_{x}(\mu,\sigma^{2})&=\frac{1}{\sqrt{\pi}}\int^{\infty}_{\frac{\alpha-\mu}{\sqrt{2}\sigma}}e^{-x^2}dx\\
&=\frac{1}{2}\left\{1-\mathrm{erf}\left(\frac{\alpha-\mu}{\sqrt{2}\sigma}\right)\right\},
\end{split}
\]
where $\displaystyle \mathrm{erf}(x)=\frac{2}{\sqrt{\pi}}\int^{x}_{0}e^{-t^2}dt$. Then we can calculate $\alpha$ by using this formula: $\displaystyle \alpha=\sqrt{2}\sigma~\mathrm{erf}^{-1}(0.98)+\mu=\frac{\sqrt{2}kT}{m\gamma}\left(t-\frac{2}{\gamma}(1-e^{-\gamma t})+\frac{1}{2\gamma}(1-e^{-2\gamma t})\right)~\mathrm{erf}^{-1}(0.98)+U\frac{1-e^{-\gamma t}}{\gamma}$.

When $P(x,t)$ is a Cauchy distribution $\displaystyle P(x,t)=\frac{1}{\pi}\frac{\Gamma}{(x-x_{0})^{2}+\Gamma^{2}}$, we have
\[
\displaystyle 
\begin{split}
\int^{\infty}_{\alpha}\frac{1}{\pi}\frac{\Gamma}{(x-x_{0})^{2}+\Gamma^{2}}dx&=\frac{1}{\pi}\int^{\infty}_{\frac{\alpha-x_{0}}{\Gamma}}\frac{1}{x^{2}+1}dx\\
&=\frac{1}{2}-\frac{1}{\pi}\tan^{-1}\left(\frac{\alpha-x_{0}}{\Gamma} \right),
\end{split}
\]

where $\Gamma$ is given by the formula: $\displaystyle \Gamma=\tilde{\Gamma}=\frac{1-e^{-\gamma t}}{\gamma}\Gamma$. Then we can calculate $\alpha$ by using this formula: $\displaystyle \alpha=\gamma\frac{1-e^{-\gamma t}}{\gamma}~\tan(0.49\pi)+U\frac{1-e^{-\gamma t}}{\gamma}$, where $\displaystyle \Gamma=U\frac{\tan 10^{\circ}}{1-\tan 10^{\circ}}$.

Table \ref{0.01} shows the distance $\alpha$ for that $1\%$ of aerosols diffuse: $\displaystyle \int^{\infty}_{\alpha}P(x,t)dx=0.01$. The result indicates that aerosols can diffuse farther when their initial distribution of velocity is Cauchy.

\begin{table}[H]
\begin{center}
\caption{The distance $\alpha$ m for that $1\%$ of aerosols diffuse}
\begin{tabular}{c||c|c}
\hline
$t~\mathrm{s}$&(Gaussian)~m&(Cauchy)~m\\ \hline
0.05&5.22&16.1\\
~0.1&8.71&22.8\\
0.15&11.0&23.9\\
~0.2&12.6&38.7\\
0.25&13.6&41.9\\

\end{tabular}
\label{0.01}
\end{center}
\end{table}

\section{Conclusion and Future Works}
In this paper, we investigate the kinetics of aerosols generated  by sneezes from the viewpoint of stochastic processes and then compare with the mean free path.  In deriving the probability distribution for position from the Langevin equation for velocity, we show that when the initial distribution is Cauchy, superdiffusion occurs immediately after emission. It means that aerosols can spread very widely in a short time, which is consistent with the fact that COVID-19 pandemic was spread to all over the world in a short period in 2020-2021. 
By using the results from the calculation using Cauchy distribution as an initial velocity distribution, we find that the required distance to prevent aerosol infection is about 42 m. Alternatively, using Gaussian distribution as an initial velocity distribution, our analytical estimation based on non-equilibrium statistical mechanics shows that the required distance is about 14 m, which is even greater than the several meter-scale (5 m) that has been previously reported in the world \cite{aeromove}. This result also supports the idea that either 1 or 2 m of which WHO and many national public health
agencies recommend maintaining physical distances is not sufficient to protect against aerosols that travel beyond this range \cite{Chia}. The results also indicate that the required distance is sufficiently larger than that of the mean free path and that the larger aerosol particles cause the farther dispersion. As a future study, how this non-equilibrium statistical physics approach of the study is consistent with the results obtained by fluid dynamics approach \cite{Bourouiba} should be investigated. 

\section*{Acknowledgement}
 The authors thank Shinji Kakinaka, Dr.  Atsushi Iwasaki and Minghui Kao for their comments on the manuscript. The authors acknowledges the project ``Study of Wireless Power Transmission Coloring for 5G Electricity'' of {\it MinnaDenryoku, Inc. } for financial support in part.

\end{document}